\documentclass[12pt]{article}
\usepackage{amsmath}
\usepackage{longtable}
\usepackage{mathtools}
\usepackage{graphicx,psfrag,epsf}
\usepackage{enumerate}
\usepackage{url} 
\usepackage{lscape}
\usepackage{xcolor}

\begin{document}

\def\spacingset#1{\renewcommand{\baselinestretch}%
{#1}\small\normalsize} \spacingset{1}
\newcommand{\ve}[1]{{\mbox{\boldmath ${#1}$}}}


  \title{\bf Novel Bayesian method for simultaneous detection of activation signatures and background connectivity for task fMRI data}
  \author{Michelle F. Miranda
  \\
    Department of Mathematics and Statistics University of Victoria\\
    and \\
    Jeffrey S. Morris \\
    Division of Biostatistics, Department of Biostatistics, Epidemiology and Informatics\\ Perelman School of Medicine, University of Pennsylvania}
  \maketitle

\bigskip
\begin{abstract}
In this paper, we introduce a new Bayesian approach for analyzing task fMRI data that simultaneously detects activation signatures and background connectivity.  Our modeling involves a new hybrid tensor spatial-temporal basis strategy that enables scalable computing yet captures nearby and distant intervoxel correlation and long-memory temporal correlation.  The spatial basis involves a composite hybrid transform with two levels: the first accounts for within-ROI correlation, and second between-ROI distant correlation.  We demonstrate in simulations how our basis space regression modeling strategy increases sensitivity for identifying activation signatures, partly driven by the induced background connectivity that itself can be summarized to reveal biological insights.  This strategy leads to computationally scalable fully Bayesian inference at the voxel or ROI level that adjusts for multiple testing.  We apply this model to Human Connectome Project data to reveal insights into brain activation patterns and background connectivity related to working memory tasks.

\end{abstract}

\noindent%
\vfill

\newpage
\spacingset{1.45} 
\section{Introduction}
\label{sec:intro}

The emergence of neuroimaging techniques has contributed significantly to our understanding of the human brain structure, function, and connectivity and their relationships to behavior. Magnetic resonance images (MRIs) are non-invasive and have high spatial resolution  compared to earlier imaging techniques such as positron emission tomography, that relies on the injection of positron-emitting radiopharmaceuticals, and electroencephalography, that generates an image with low spatial resolution. As a result, they have become the most popular choice in large neuroimaging studies. Together, MRIs such as diffusion tensor images (DTI) and functional MRI (fMRI) can provide an effective approach to analyzing brain circuitry and identifying locations of brain activation. In this paper, we describe a novel Bayesian methodology to analyze task-based fMRI data while simultaneously pursuing these two goals.

\subsection{Task-based fMRI}
Task-based functional magnetic resonance imaging (fMRI) studies are a powerful tool to understand human sensory, cognitive, and emotional processes. In the Human Connectome Project (HCP), a core battery of tasks were designed to not only identify and characterize the functionally distinct nodes in the human brain, but also to relate signatures of activation magnitude or location in key network nodes. These results can in turn be used to validate the outcomes of the connectivity analyses on resting state fMRI (R-fMRI), resting state MEG (R-MEG), and diffusion data (\cite{Barch2013}). Among the broad range of task domains available in the HCP study, we focus here on data from the Working Memory task.

The term ``Working Memory'' refers to the brain process of storing and manipulating information involved in the performance of complex cognitive tasks  (\cite{Baddeley2010}). To optimally perform a task the brain will enter a task state and it needs to maintain it throughout the task. It is hypothesized that this is done by brain modulation of task-dependent connection patterns. \cite{Elkhetali2019} use the term ``background connectivity'' for the task-dependent modulations that are due to variations in ongoing brain activity instead of stimulus-driven activity. To fully understand brain functionality during tasks, it is crucial to look at ongoing activity in addition to stimulus-driven activity. 

In the HCP, the Working Memory data was acquired in two runs, one with a right-to-left and the other with a left-to-right phase encoding. Within each run, half of the blocks used a 2-back working memory task and half used a 0-back working memory task. Additionally, a component consisting of four categories of pictures (faces, places, tools, and body parts) was embedded within the working memory task. Therefore each of the two runs contains eight task blocks. To find activation signatures, we can either ignore stimulus type and focus on memory load comparisons, or we can collapse across memory load and focus on  stimulus comparisons between the  image categories to identify temporal, occipital, and parietal regions. There is evidence that the four categories of images engage distinct cortical regions, with differences between ``faces'' and other categories localized mainly on the fusiform face area and occipital gyrus (\cite{Barch2013}). Background connectivity usually follows from first computing the residuals after model estimation and then estimating correlations from these computed residuals. Therefore, a unified modelling approach that accounts for the complex spatial and temporal structures and jointly estimates background connectivity will not only improve the power to correctly identify important activation signatures but will also provide useful inference on background connectivity patterns.

\subsection{Task-based fMRI models}

A critical challenge is to account for both spatial and temporal dependencies from the complex structure of the brain. The simplest models for fMRI treat voxels as independent units during estimation, and either ignore time dependencies or assume the noise follows an auto-regressive process. Particularly in Bayesian models, a common approach is to assume an autoregressive structure of order $q$ with a prior distribution on the AR coefficients (\cite{Lee2014, Penny2003, Woolrich2004}).
Following estimation, inference is often based on statistical parametric maps and significance is obtained based on the Gaussian Field theory. The theory is heavily dependent on the smoothness properties of the image which is obtained by using a Gaussian kernel with the full-width-half-max (FWHM) in the range of 8-16mm as a pre-processing step (\cite{Friston2007}). However, different amount of smoothing can lead to different results (\cite{Liu2017}). Specifically, misspecification of the FWHM that can blur activations or leave unnecessary noise in the data, leading to a suboptimal approach (\cite{Flandin2007}).

Bayesian spatiotemporal models are particularly attractive in fMRI applications because they can easily incorporate spatial correlation by either assuming a spatial prior directly on the regression coefficients $b_j(v)$ in Model \eqref{eq:voxelwise} or by assuming a spatial prior on the activation characteristic parameters of the voxels. These priors are usually either an Ising prior or a Gaussian Markov random field prior  (\cite{Gossl2001,Penny2005, Quiros2010,Lee2014,Harrison2010}), which are able to capture nearby voxel correlations. As for temporal structure, all these models assume either an independent noise structure for $e_t(v)$ in Model \eqref{eq:voxelwise} or an autoregressive structure.

More flexible  models for the noise in fMRI data include the assumption of  long memory covariance structures.  Biological data, such as functional magnetic resonance images of the human brain, often demonstrate scale invariant or fractal properties. Fractal time series typically have long-range autocorrelations (long memory) in time (\cite{Bullmore2004}). This feature has motivated a large literature of discrete wavelet transforms (DWT) in fMRI (\cite{Fadili2002, Sanyal2012, Zhang2014, Zhang2016}). DWTs are particularly attractive because the correlation between wavelet coefficients will generally be small even if the data are highly autocorrelated (\cite{Bullmore2004}).  \cite{Fan2003} showed that the DWT has optimally decorrelating properties for a wide class of signals with 1/f-like or long memory characteristics.

Following these principles, \cite{Zhang2014} assume a long-memory correlation structure for the errors and project the BOLD time series using a discrete wavelet transform. The authors account for spatial correlation among nearby voxels by using a Markov random field (MRF) prior on the parameters guiding the selection of the activated voxels in a 2D slice of the brain and impose a Dirichlet Process prior on the parameters of the long memory process.  One major limitation of their approach is that their model is fitted in each brain slice and does not account for the complex 3D characteristics of the brain volume. Moreover, their spatial model assumptions in the voxel domain increase computational time and do not scale up to multiple subjects.

In this paper, we introduce a novel Bayesian method for simultaneous learning of activation signatures and background connectivity.  Our method is based on a novel spatial-temporal tensor basis modeling strategy that captures key aspects of the complex structure of the fMRI data in an adaptive way. Our spatial basis approach models voxel level data yet incorporates biological information such as contained in known anatomical regions through prespecified Regions of Interest (ROI).  It involves a composite hybrid basis strategy that first discovers intra-ROI voxel-level structure, and then in a second level models inter-ROI structure that can capture distant correlation from background connectivity. In this way, our strategy inherits the interpretability of ROI-based modeling strategies and the flexibility of voxel-based modeling approaches.  We consider time dependencies in the BOLD time series by assuming a long memory process in the space of reduced spatial features, efficiently accounting for this structure through a temporal wavelet basis transform.  We fit a fully Bayesian version of this model via MCMC, that yields Bayesian inference that adjusts for the inherent multiple testing problems.  The novel methodology allows for estimation of background (residual) connectivity between ROIs that are implicit to the modeling and helps improve sensitivity for detection of activation signatures.

The contributions of the proposed model are many fold. First, our model appears to have a better detection power than the state-of-the-art methods and other spatial basis approaches, which is obtained by borrowing both spatial and temporal information through the carefully design basis approach. Second, the composite-hybrid basis approach provides a sparse spatial representation of the brain while yielding full Bayesian inference at the voxel and ROI level with incredible computational speed. Third, the model allows for full Bayesian estimation of the background connectivity, which not only increases the power to estimate signature locations but also allow scientists to gain insights of the underlying brain function necessary to maintain a task state.  

The rest of the paper is organized as follows. In Section \ref{sec:meth} we introduce our model and novel tensor basis construction strategy, describe our Bayesian model-fitting approach, explain how to estimate inter-ROI background connectivity, and present our Bayesian inferential approach for detecting activation signatures that adjusts for multiple testing.  In Section \ref{sec:sim} we compare our proposed adaptive composite-hybrid basis modeling strategy with various alternative spatial basis approaches and mass univariate voxel-level methods using simulated data. In Section \ref{sec:app} we apply our method to the Working Memory Data from the Human Connectome Project and describe what biological insights it reveals. Finally, in Section \ref{sec:discus}, we present some concluding remarks.

\section{Methods}
\label{sec:meth}

\subsection{Functional Regression Model for fMRI Activation}
The fMRI measurements comprise an indirect and non-invasive measure of brain activity based on the Blood Oxygen Level Dependent (BOLD) contrast (\cite{Ogawa1990}). For each voxel $v=1,\ldots,N_v$ in a volumetric image, the fMRI response function is a time series with $T$ time points of BOLD contrast values that can be modeled as

\begin{equation}
\label{eq:voxelwise}
  y_t(v)= \sum_{j=1}^p b_j(v)(s_j*h)(t)+e_t(v),  
\end{equation}

 \noindent where  $s_j(t)$ is a indicator function of stimulus $j$ at time $t$, $h(t)$ models the haemodynamic response of a neural event at time $t$ (\cite{lindquist2008}), and $e_t(v)$ is a measurement noise. The coefficients $b_j(v)$ characterize the relationship of brain activity during stimulus $j$. The primary goal of activation studies is to identify which voxels are differentially activated by specified stimuli.

Model \eqref{eq:voxelwise} can be written in matrix notation as

\begin{equation}
\label{eq:originaldata}
  \ve Y=  \ve X \ve B+ \ve E,  
\end{equation}

\noindent where $\ve Y$ is a $T \times N_v$ matrix of BOLD response measurements, $\ve X$ is a $T \times P$ design matrix formed by the convolution of $s$ and $h$,  $\ve B$ is a $P \times N_v$ matrix of regression coefficients, and $\ve E$ is a matrix of residual errors assumed to follow a matrix normal distribution $\ve E\sim NM_{T \times N_v}(\ve 0, \ve U, \ve L)$, with the structure of the temporal correlation matrix $\ve U$ and spatial correlation matrix $ \ve L$ being determined by the basis modeling strategy described in Section \ref{sec:comp-hybrid}. Correlation among the residual errors in an activation model like this has been called \textit{background connectivity} (\cite{Elkhetali2019}).

\subsection{Adaptive Composite-Hybrid  Modeling Approach}
\label{sec:comp-hybrid}

 We will use a functional regression 
 strategy to fit this model, using basis functions to represent the 4d spatiotemporal fMRI data and model components.  This strategy regularizes the regression coefficients across the volumetric space and accounts for intervoxel spatial correlation in the residuals, which in principle can produce more efficient estimation and increased power for detecting activated brain regions, while also providing estimates of background connectivity and inducing dimension reduction that will substantially speed calculations to enable a fully Bayesian modeling approach.  In principle, this approach can yield greater precision and power than the standard mass univariate voxel-level approaches that are commonly used by fMRI practitioners.
  
We will use a tensor product of temporal and spatial basis functions, with $\Xi=\{\Theta,\Upsilon\}$ representing the set of basis functions with $\Theta$ a temporal and $\Upsilon$ a spatial basis defined on the temporal and volumetric spaces, respectively.  Given these bases, we will transform the spatiotemporal data and model \eqref{eq:originaldata} into the corresponding tensor basis space, which can be represented as: 
\begin{eqnarray}
    \Theta\ve Y \Upsilon=  \Theta \ve X \ve B \Upsilon+ \Theta\ve E\Upsilon.  \nonumber  \\
    \ve Y^* = \ve X^* \ve B^* + \ve E^*, 
        \label{eq:projectedmodel}
\end{eqnarray}
\noindent where the rows of $\ve Y^*$, $\ve X^*$, and $\ve E^*$ index the temporal basis coefficients and columns of $\ve Y^*$, $\ve B^*$, and $\ve E^*$ index the spatial basis coefficients for the respective model components. Specific modeling assumptions are described below.

In this \textit{tensor basis space model}, we can efficiently model the  columns using fast independent regressions with independent rows, with the basis space modeling automatically accounting for the complex inherent spatiotemporal correlation structure of the data (\cite{Morris2015}). As we will describe next, we will introduce for the spatial basis $\Upsilon$ a novel \textit{composite-hybrid} basis strategy to empirically estimate both nearby (local) and distant (global) spatial dependencies, and for the temporal basis $\Theta$ will use a wavelet to capture long memory temporal dependencies in the BOLD time series.

 \subsubsection{Spatial basis $\Upsilon$}
 \label{model:spatialbasis}
 
 Various alternatives could be used for the spatial basis $\Upsilon$.  One simple idea that might seem natural to some is to use principal component (PC) basis functions empirically estimated from the voxel-level data, a strategy we denote \textit{global spatial basis (GSB)}.  While in principle flexible and able to empirically estimate the spatial dependence structure, as we will demonstrate this strategy does not work well because there is typically not enough data to reliably estimate these global PCs.
 
 Another simple option some might consider would be to sum voxels within pre-specified regions of interest (ROI), e.g. from anatomical brain maps, and detect differential activation at the ROI level.  This is equivalent to using basis functions with loadings of unity within the respective ROIs and zero elsewhere.  While incorporating biological information and computationally efficient, this strategy can lose information if the differentially activated regions do not correspond to the ROIs, and does not really provide a voxel-level analysis.  
 
We will introduce a novel strategy that can be viewed as a compromise between these two extremes.  It involves a multi-level approach resulting in what we dub \textit{composite-hybrid} basis functions: first using ROI-specific PCA to identify modes of variability within the ROIs, and then applying a second level PCA to these \textit{local} PC scores to capture distant correlation across the ROIs.  This enables a full voxel-level analysis that flexibly accounts for the spatial correlation structure yet incorporates the biological information inherent in the ROIs for interpretability and enhanced dimension reduction.   
 
  \paragraph{First level - Intra ROI basis representation.}
 
Assume we divide the brain volume $\ve Y$ into $K$ pre-specified clusters $C_1,\ldots,C_K$, e.g. ROIs.  Let $Y^{(k)}_t(v)= Y_t(v)\mbox{I}_{\{v \in C_k\}}$, where $C_k$ indicates the $k$th cluster, and its matrix version $\ve Y^{(k)}$ be of size $T \times n_k$. If the clusters represent a perfect partition of the brain volume,  then the columns of $\tilde{\ve Y}=\{\ve Y^{(1)},\ldots,\ve Y^{(K)}\}$ are a reordered version of the columns of $\ve Y$ and therefore $\sum_{k=1}^{K}n_k=N_v$. 
Other times, ROIs are only defined on what are deemed to be biologically relevant brain regions, in which case $\tilde{\ve Y}$ may only comprise a subset of the brain volumetric space with $\sum_{k=1}^{K}n_k<N_v$. 

This first level captures local features representing modes of variability within the ROI by projecting data of each ROI, $\ve{Y}^{(k)}$, into the space spanned by its principal components. The new feature matrix for the $k$th ROI is given by
\begin{equation}
\label{eq_trans1}
\ve Y^L_k=\ve{Y}^{(k)} \Phi^{(k)}=Y_t(v)\Phi^{(k)}\mbox{1}_{\{v \in C_k\}},\end{equation}

\noindent where the columns of $\Phi^{(k)}$ are the orthonormal eigenvectors of ${\ve{Y}^{(k)}}^{T}\ve{Y}^{(k)}$, also called principal components, and therefore $\ve{Y}^{(k)}= \ve Y^L_k{\Phi^{(k)}}'$. The function $\mbox{1}_{A}$ represents the indicator function of the event $A$.

Assume that the columns of ${\Phi}^{(k)}$ are ordered in descending order of the corresponding eigenvalues $\{\lambda^{(k)}_l, l=1,\ldots, L\}$. For dimensionality reduction and regularization, the number of principal components in $\Phi^{(k)}$, $p_k$, can be chosen such that $p_k=\min_L(\mbox{Var}{\Phi}^{(k)}_L\geq a^L_k)$, with $\mbox{Var}{\Phi}^{(k)}_L=\sum_{l=1}^L\lambda^{(k)}_{l}/\sum_{s=1}^{S_k} \lambda^{(k)}_{s}$ and $S_k$  the total number of nonzero eigenvalues. The scalar $a^L_k \in (0,1)$ indicates the proportion of the variance of ${\tilde{\ve Y}_k}$ that is explained by $\ve Y^L_k$. 

Let $\ve Y^L=[{\ve Y^L_1},\ldots,{\ve Y^L_K]}=\tilde{\ve Y}\Phi$ be a matrix of size $T \times \sum_{k=1}^K p_k$ of local features, where
\[
    {\ve \Phi}=\left(
    \begin{array}{ccc}
    \ve \Phi^{(1)} && 0\\
    & \ddots &\\
    0 && \Phi^{(K)}           \\
    \end{array}
    \right)
\]  

\noindent is the local spatial projection matrix. Then, $\ve Y^L$ lies in the {\itshape space of local features}.

 \paragraph{Second level - Inter ROIs basis representation.}  To capture long-range spatial correlations, we project the local feature matrix, $\ve Y^L$, into the space spanned by its principal components. This projection accounts for inter-ROI dependencies and provides additional dimension reduction. The resulting matrix contains the global-spatial features of the brain data. We write
\begin{equation}
\label{eq_trans2}
\ve Y^{G}=\ve Y^{L}\ve\Psi=\tilde{\ve Y} \ve\Phi \ve\Psi, \end{equation}

\noindent where the columns of $\ve \Psi$ are the orthonormal eigenvectors of $({\ve{Y}^L})^{T}(\ve{Y}^L)$. Therefore $\ve{Y}^L= \ve Y^{G}{\ve \Psi'}$ and $\tilde{\ve Y}= \ve Y^{G}{\ve \Psi' \Phi'}$.
Similarly to the previous section, the number of components in $\ve \Psi$, $S$, is chosen such that the
$S=\min_M(\mbox{Var}{\Psi}_M\geq a^G_k)$, where $\mbox{Var}{\Psi}_M=\sum_{m=1}^M \lambda^G_{m}/\sum_{r=1}^{R}\lambda^G_{s}$, $\{\lambda^G_m; m=1,\ldots, R\}$ are the eigenvalues in descending order, with $R$ the number of nonzero eigenvalues. Note that the size of the global feature matrix $\ve Y^{G}$ is $T \times S$, meaning that the number of BOLD times series that characterize the fMRI experiment in the voxel space was $N_v$ but in the spatial basis space is $S$. Typically, $S \ll N_v$, and we have found this reduces dimensionality from near a million down to 300-400.

We call this a \textit{composite-hybrid spatial basis} strategy since it effectively uses a spatial basis that is a composition of the local and global basis functions, $\Upsilon=\ve\Phi\ve\Psi$.  This strategy is able to capture global spatial correlations inherent to background connectivity like the global PCA while being feasibly computed from the available data, and incorporates the biological information inherent to the ROIs yet providing a full voxel-level analysis.

 \subsubsection{Accounting for temporal correlation via temporal basis functions} \label{model:temporalbasis}
The matrix of global features $ \ve Y^{G}$ has orthogonal columns corresponding to a series of $S$ BOLD time series in the spatial basis space, but the rows are temporally correlated.  To account for temporal correlation we will use a wavelet representation that can induce a long memory process that accouns for the temporal correlation. 

Let $\Theta=\ve W$, where $\ve W$ is a $T^W \times T$  is a matrix representing a discrete wavelet transform (DWT) for a chosen wavelet basis of size $T^W$ at $M$ resolution levels with $N_m$ coefficients at each level $m=1,\ldots,M$, with $T^W=\sum_{m=1}^M N_m$. 
Let $\ve Y^*$ be a $T^W \times S$ matrix with rows transformed into the wavelet space, with

\begin{equation*}
\label{eq:wavelettrans}
   \ve Y^*=\ve W (\ve Y^{G})= \ve W \tilde{\ve Y} \ve\Phi \ve\Psi.
\end{equation*}

\noindent We can then write Model \eqref{eq:projectedmodel} as

\begin{equation}
    \label{eq:projectedfinalmodel}
    \ve Y^{*}=  \ve X^* \ve B^*+ \ve E^*,    
\end{equation}
\noindent where $\ve X^*= \ve W\ve X$ is a $T^W \times P$ design matrix, $\ve B^*=\ve B\ve\Phi \ve\Psi$ a $P \times S$ matrix of basis space regression coefficients, and $\ve E^*=\ve W \ve E \ve\Phi \ve\Psi$ a $T^W \times S$ matrix of residuals.

We can represent the components of Model \ref{eq:projectedfinalmodel} in vectorize form. Let

\[\mbox{vec}(\ve Y^*)= (\ve\Psi^T \ve\Phi^T \otimes \ve W) \mbox{vec}(\tilde{\ve Y}) \]
and
\[\mbox{vec}(\ve E^*)= (\ve\Psi^T \ve\Phi^T \otimes \ve W) \mbox{vec}(\ve E), \]
we can rewrite Model \ref{eq:projectedfinalmodel} as
\begin{equation}
 \mbox{vec}(\ve Y^*)=( \ve I_S \otimes \ve X^*)( \ve I_S \otimes \ve B^*)+  \mbox{vec}(\ve E^*)
\end{equation}

\noindent We will assume  $\mbox{vec}(\ve E^*) \sim N(0,\Sigma^W)$, with $\Sigma^W=\mbox{diag}(\Sigma^W_1,\ldots,\Sigma^W_S)$ being a block diagonal matrix. Each \{$\Sigma^W_s, s=1,\ldots,S$\} is a $T^W \times T^W$ diagonal matrix with elements $\psi_s\sigma^2_{mn}=\psi_s(2^{\alpha_s})^{-m}$, indicating the variance of the $n$th wavelet coefficient at the $m$th scale. The parameter $\psi_s$ is called the \textit{innovation variance} and ${\alpha_s} \in (0,1)$ the \textit{long memory parameter}. 

These simpler assumptions in the wavelet space have been used to account for possible long memory structure of the BOLD time series
 (\cite{Zhang2014,Zhang2016}). We specify a prior $\psi_s \sim \mbox{Inverse Gamma}(a_0,b_0)$ and estimate $\alpha_s$ by the method of moments, as described in the Supplementary Material.
 
 We assume a prior distribution on the basis space regression coefficient matrix. Let $\ve b_{s}^* \sim \mbox{N}(0,\tau_s^2)$ be the $s$th column of $\ve B^*$.  We assume $\tau_s^2=k_v(\ve X^{*T}({\Sigma^W_s})^{-1}\ve X^*)^{-1}$, a function of the unknown variance components $\Sigma_s^W$, and  choose the constant $k_v$ between 10 and 100 to induce an appropriate level of shrinkage. In functional regression, regularization of functional coefficients is accomplished by truncation, shrinkage, or sparsity in the basis space (\cite{Morris2015}).  The combination of truncating the local and global PC basis and the shrinkage induced by this prior will spatially regularize the volumetric space regression coefficients $b_j(\nu)$ across voxels $\nu$, with the $\tau_s^2$ affecting the degree of regularization.

\subsubsection{Induced Background Correlations}
\label{subsubsec:indcorr}
The block diagonal matrix $\Sigma^W$, after inverse basis transformations, induce a background correlation matrix in the space of spatial features in which $\ve Y^{G}$ lies, in the space of local features in which $\ve Y^{L}$ lies, and in the original volumetric data space. The corresponding spatial covariance matrix in the volumetric space makes it possible to estimate background connectivity.
For interpretability, we focus on estimating the background correlations between ROIs induced by our empirically determined hybrid spatial basis functions.

 Let $\Sigma_{ROI}$ be a $\sum_{k=1}^K p_k \times \sum_{k=1}^K p_k $ of induced background correlation in the ROI space is given by $\Sigma_{ROI}=\Psi \Sigma_{\theta} \Psi'$, where $\Sigma_{\theta}= \mbox{diag}(\theta_1,\ldots,\theta_S)$ and $\theta_s=\mbox{mode}(\mbox{diag}(\ve W^T\Sigma^W_s \ve W))$. A detailed derivation of this matrix can be found in the Supplementary Material. This matrix can be investigated to assess the background connectivity at the ROI level.

Thus, the composite-hybrid structure allow for background correlations to be estimated, which can provide further insights into ongoing brain activity. It is important to notice that the connectivity estimation is only possible because of the second-level global basis estimation inherent in the composite-hybrid bases $\Upsilon=\Phi\Psi$, since if only the local basis from the first step were used, i.e. $\Upsilon=\Phi$ which we call \textit{Local Spatial Basis} (LBS), then $\Sigma_{ROI}=\Sigma_{\theta}$ which would be a diagonal matrix and there would be no estimated inter-ROI background connectivity estimates. 

\subsection{Estimation}
\label{subsec:estimation}
Following are the steps for fitting our model -- details are provided in the Supplementary Material. We obtain posterior samples of $\ve B^*$ in Equation \eqref{eq:projectedfinalmodel} then use the inverse DWT (IDWT) and the properties of the orthogonal basis described on Section \ref{model:spatialbasis} to obtain posterior samples of $\ve B$ (Equation \eqref{eq:originaldata}) in the original data space. We use the posterior samples of $\ve B$ to determine activation signatures, and the posterior samples of $\psi_s, s=1,\ldots, S$ to estimate the background connectivity as detailed in sections \ref{subsec:simbas} and \ref{subsec:RV}, respectively.

\subsection{Bayesian determination of activation signatures}
\label{subsec:simbas}
From posterior samples, we can construct joint credible bands over the entire brain volume using the approach of \cite{Ruppert2003} as done in other applications of Bayesian functional regression \cite{Meyer2015}.  These joint bands effectively account for multiple testing across the entire voxel space, so we will use these to flag activation signatures based on voxels for which the $100(1-\alpha)\%$ joint credible bands exclude 0.

Let ${\ve C}=\sum_a d_a B_a$ be a contrast in the data space representing a desired comparison of interest, e.g. comparing working memory tasks. Let $M$ be the number of MCMC samples for the contrast, after burn-in and thinning. A joint $100(1-\alpha)\%$ credible interval for $\ve C$ is given by

\[I_{\alpha}(v)=\hat{C}(v) \pm q_{(1-\alpha)}[\widehat{\mbox{std}}(\hat{C}(v))],
\]

where $\hat{C}(v)$ is the contrast posterior mean, and $\widehat{\mbox{std}}(\hat{C}(v))$ is the posterior standard deviation. The $1-\alpha$ quantile, $q_{(1-\alpha)}$, is taken over $M$ from the quantity

\[z^{(m)}=\max_{v\in \mathcal{V}}\left|\frac{C^{(m)}(v)-\hat{C}(v) }{\widehat{\mbox{std}}(\hat{C}(v))} \right|.\]

Following \cite{Meyer2015}, if desired, these joint credible bands can be inverted to obtain simultaneous band scores, or \textit{SimBas}, for each voxel that determines the minimum $\alpha$ for which the $100(1-\alpha)\%$ credible interval excludes zero. Denote by $P_{SimBas}$ the minimum $\alpha$ at which each interval excludes zero ($P_{SimBas}=\min\{\alpha: 0 \notin I_{\alpha}(v)\}$). $P_{SimBas}$ can be directly computed by
\[P_{SimBas}(v)=\frac{1}{M}\sum_{m=1}^M1\left\{\left|\frac{\hat{C}(v)}{\widehat{\mbox{std}}(\hat{C}(v))}\right|\leq z^{(m)}\right\}.\]

\subsection{Estimating Background Connectivity in the ROI space}
\label{subsec:RV}
 To estimate the background connectivity between clusters, we use the multi-scale approach proposed by \cite{Ting2020}. We estimate the global background connectivity (between ROIs)  across regions using the square root of the RV coefficient as a single-value cross-dependence measure for the linear dependence between components of different ROIs (\cite{Ting2020,Escoufier1973}).
 
 Let the cluster $C_{r_jr_k} = C_{r_j}\cup C_{r_k}$ be the union of the voxels in regions $r_j$ and $r_k$. Let $\Sigma_{ROI}^{r_jr_k}$ be the background correlation matrix associated with the local features in $C_{r_jr_k}$. The RV coefficient between features in ROIs $r_j$ and $r_k$ is defined by
\[ \mbox{RV}_{r_jr_k}=\frac{\mbox{trace}(\ve M_{r_jr_k}\ve M_{r_kr_j}')}{\sqrt{(n_in_j)}},\]

\noindent where $\ve M_{r_jr_k}=(\Sigma_{ROI}^{r_j})^{-1/2}\Sigma_{ROI}^{r_jr_k}(\Sigma_{ROI}^{r_k})^{-1/2}.$ Details on how to obtain $\Sigma_{ROI}$ are in the Supplementary Material.

\section{Simulation Studies}
\label{sec:sim}
The purpose of this set of simulation studies is to show that the composite-hybrid basis model approach not only is able to capture true activation regions, but also to estimate the brain background connectivity. Additionally, we perform a null simulation that mimics resting state data to investigate the false positive rates of the composite-hybrid basis model. To make our simulations more realistic, we based our simulation upon a real dataset: the working memory task data from Section \ref{sec:app}. To generate the data, we consider a long-range spatial and 
 auto-regressive in time error to model the temporal correlations. To generate the long-term error structure, we first generate a short-range spatial auto-regressive error as described below.
\paragraph{Short-range spatial and autoregressive in time correlation structure.}
Let $ v$ be a voxel in the three-dimensional space, then ${E}^{(1)}_{t}(v) = \sum _{\lVert v'-v \rVert_1 \leq 1} {E}^*_{t}(v')/m_v$, where $\lVert . \rVert$ is the $L_1$ norm of a vector and $m_v$ is the number of locations in the set $\{\lVert v'-v \rVert_1 \leq 1\}$. If independence is assumed in the temporal domain, then ${E}^*_{t}(v)\sim \mbox{N}(0,1)$, otherwise we assume an auto-regressive of order 2 and ${E}^*_{t+2}(v)=0.9{E}^*_{t+1}(v)-0.8{E}^*_{t}(v)+\epsilon_{t+2}(v)$, with $\epsilon_{t}(v)\sim \mbox{N}(0,0.5)$.

\paragraph{Long-range spatial and auto-regressive in time correlation structure.}
Let $ v=(v_1,v_2,v_3)$ be a voxel in the three-dimensional space, then $E^{(2)}_{t}(v) = 2 \sin(\pi v_1/10)e_{t,1}+ 2 \cos(\pi v_2/10)e_{t,2}+2 \sin(\pi v_3/5)e_{t,3}+ E^{(1)}_{t}(v)$, where $e_{tk}$ for $k=1,\ldots,3$ were independently generated from a N$(0,0.5)$ generator and $E^{(1)}_{t}(v)$ is the short-range spatial and auto-regressive in time error.

\subsection{Activation and background connectivity}

\begin{figure}
    \centering
    \includegraphics[width=0.85\textwidth]{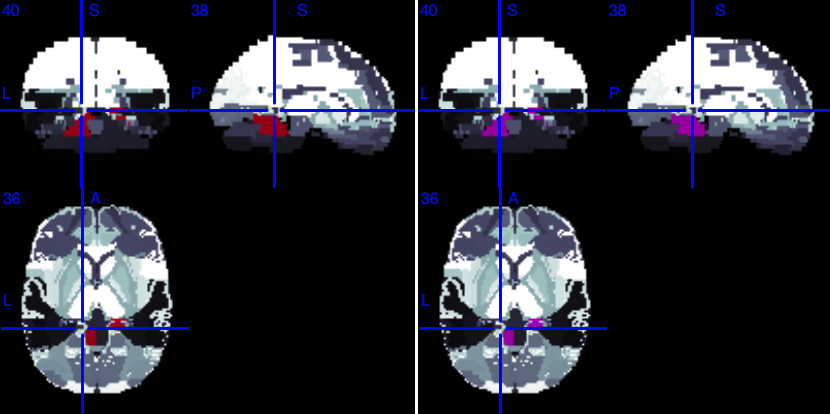}
    \caption{Simulation Results. The left panel shows the contrast $\ve B_1 (v)-\ve B_2 (v)$ used to simulate the data. The right panel shows brain locations where there is a difference in activation between stimulus 1 and 2, based on the joint bands described in Section \ref{subsec:simbas}. }
    \label{fig:sim1}
\end{figure}

In this example, we simulate data to mimic brain activation in two distinct regions of the brain during a simple task. We assume that the BOLD signal at time  $t=1,\ldots,100$, and voxel $v \in R^{32 \times 32 \times 25}$ is modeled by
\begin{equation}
    Y_{t}(v)= \bar{Y}_C(v)+\kappa_0{\ve X_t}{\ve B}(v)+\kappa_1U_{t}(v)+\kappa_2E_{t}(v), \nonumber
\end{equation}

 \noindent where the covariate $\ve X=\{\ve X_t, t=1,\ldots,100\}$ is a matrix with two columns corresponding to the convolution of two different stimulus indicators and the double-gamma HRF. The coefficients $\ve B_1 (v)$ and $\ve B_2 (v)$ characterize the relationship of brain activity during each stimulus and are shown in and blue and red, respectively, on Panel (a) of Figure \ref{fig:sim1}. The quantity $\bar{Y}_C(v)$ is the average taken over all time points of a true brain signal from the example subject described in Section \ref{sec:app}. The example subject data is originally in $R^{91 \times109\times 91}$ and we took a hyperrectangle of size $32 \times 32 \times 25$ to compute $\bar{Y}_C(v)$ and to specify the clusters as defined in Section \ref{model:spatialbasis}. We then have 33 clusters given by true biological regions in the specified hyperrectangle and considering the Tailarach labels (\url{http://www.talairach.org/about.html#Labels}). These clusters are shown in Panel (a) of Figure \ref{fig:sim1} and are overlaid in the original Tailarach atlas consisting of 1024 distinc regions of interest. 
 
The noise components $U_t(v)$ and $E_t(v)$ reflect the spatial and temporal correlations in the model. We assume $U_t=e_t\Lambda$, where $\Lambda$ is an $N_r \times N_v$ indicator matrix with entry 1 if voxel $v$ for $v=1,\ldots, N_v=25600$ belongs to cluster (or ROI) $r$, for $r=1,\ldots, N_r=33$. Let $e_t \sim N(0,\Sigma_{roi})$, then $U_t \sim N(0,\Sigma_U)$ with  $\Sigma_U=\Lambda^T\Sigma_{roi} \Lambda$. 
We specify $\Sigma_{roi}$ to reflect the background correlation between ROIs. We consider $\Sigma_{roi}(1,22)=0.7$, $\Sigma_{roi}(4,5)=0.6$ and $\Sigma_{roi}(2,7)=0.8$. We assume that $E_t(v)$ follows the long-range spatial and auto-regressive in time correlation structure previously described. Finally $(\kappa_0, \kappa_1, \kappa_2)$ are constants chosen to control the signal to noise ratio. Here we choose $\kappa_0=1$, $\kappa_1=0.5$, and $\kappa_2=2.2$.

\begin{figure}
    \centering
    \includegraphics[width=\textwidth]{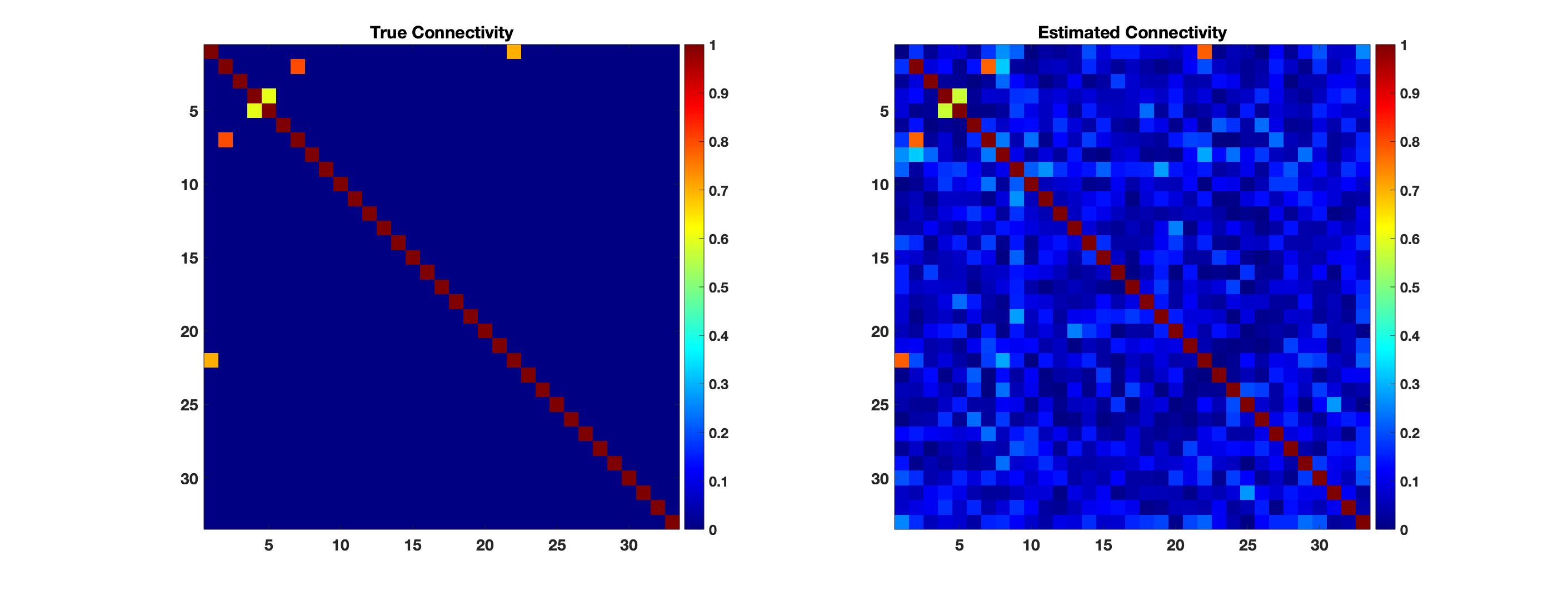}
    \caption{Simulation Results. The left panel shows the true background connectivity, $\Sigma_{roi}$, and the right panel shows the estimated background connectivity as described in Section \ref{subsec:RV}.}
    \label{fig:sim1_connec}
\end{figure}

Figures  \ref{fig:sim1} and \ref{fig:sim1_connec} show the simulation results. Figure  \ref{fig:sim1} shows brain locations where there is a difference in activation between stimulus 1 and 2, based on the joint bands described in Section \ref{subsec:simbas}.  Figure \ref{fig:sim1_connec} shows the background connectivity used to simulate the data and the estimated background connectivity based on the joint bands described in Section \ref{subsec:simbas}. Our proposed composite-hybrid model effectively capture real differences in activation regions from different stimulus and true background connectivity patterns.

We compared the following models: (i) CHSB (composite-hybrid spatial basis) with $\Upsilon=\ve\Phi\ve\Psi$; (ii) LSB (local spatial basis) with $\Upsilon=\ve\Phi$; (iii) GSB (global spatial basis) with $\Upsilon=\ve\Psi$, the principal components approach for the spatial domain; and (iv) NBM (non-basis model) where we considered spatial independence across voxel. For all models, we considered the temporal correlation as described in Section \ref{model:temporalbasis}. For each model, we computed the average joint bands interval width across voxels, with joint bands were computed as described in Section \ref{subsec:simbas}. Table \ref{table:sim1} shows the results. 

Comparing the no-basis approach with the three choices of spatial basis, we notice that accounting for spatial dependencies through a basis translates into tighter joint bands and a better precision.  This makes sense, since accounting for the intervoxel correlation reduces the effective number of multiple tests across voxels, thus tightening joint bands and gaining sensitivity for detecting differentially activated regions. Although none of the spatial choices showed any false-positives, the spatial basis models were able to capture true positives in a much higher rate than the no-basis approach.  Note also the improved MSE and much faster run times obtained when using the basis function modeling strategies.  While it appears our proposed CHSB has the best MSE, true positive rate, and precision, we performed more extensive simulations to evaluate these comparisons.

We replicated the simulation in Table \ref{table:sim1} one hundred times to get a more detailed comparison of the three spatial bases.  We omitted the "no basis" option since it took so long to run and was clearly inferior.  Table \ref{table:sim2} summarizes the results of these simulation in terms of MSE, RP, and average interval widths as previously described.  We omit FP since they were zero for all methods and data sets for these simulations.
Note that while the global PCA (GSB) performed reasonably well, both the local spatial basis (LSB) and our recommended compositive hybrid spatial basis (CHSB) had better MSE and RP, much tighter joint credible bands, and faster calculations.  CHSB and LSB performed similarly, both capturing all of the correct activated regions, with the joint bands for CHSB slightly tighter and run time slightly faster.  In real data applications with higher number of voxels and multiple subjects, the gains in computational speed are substantial as noted in Section \ref{sec:app}.

\begin{center}
\begin{table}
  \centering
\begin{tabular}{|c|c|c|c|c|}
\hline
Model & CHSB & LSB & GSB & NSB \\
\hline
Average Width & 0.885 & 0.893 & 1.152&1.275  \\
MSE&0.0147 & 0.0176 & 0.0161 & 0.0194\\
FP & 0 &0&0&0\\
RP & $> 0.9999$ & $> 0.9999$ &0.7426 & 0.6950\\
Time (seconds) & 17.2 & 18.2 & 39.7 & 6865.2 \\
\hline
\end{tabular}
\caption{Average joint bands interval width computed across all 25600 voxels, mean squared errors (MSE) of the contrast estimator, average rate of false-positive voxels (FP), and average rate of real positive voxels (TP). Comparisons are between the following models (i) CHSB  (composite-hybrid spatial basis) with $\Upsilon=\ve\Phi\ve\Psi$; (ii) LSB (local spatial basis) with $\Upsilon=\ve\Phi$; (iii) GSB (global spatial basis) with $\Upsilon=\ve\Psi$, the principal components approach for spatial dependencies; and (iv) NSB (no spatial basis). }
\label{table:sim1}
\end{table}
\end{center}


We repeated the comparison of the three spatial basis using the short-range spatial error described in the beginning of this section and display them in the Supplementary Material. The results confirm the findings observed in the previous paragraph. The global spatial basis model performs better under the short-range spatial noise term than under the long-range spatial noise term as expected. However, its performance is still inferior to the models involving the CHSB and the LSB.

\begin{table}[htbp]
  \centering
    \begin{tabular}{|l|ccccc|}
       \hline
    \textbf{CHSB} & Mean  & SD    & Min   & Median & Max \\
    Average Width & 0.841 & 0.026 & 0.772 & 0.841 & 0.911 \\
    MSE   & 0.015 & 0.005 & 0.006 & 0.014 & 0.027 \\
    RP    & $>0.999$     & 0     & $>0.999$     & $>0.999$     & $>0.999$ \\
        \hline
    \textbf{LSB} & Mean  & SD    & Min   & Median & Max \\
    Average Width & 0.850 & 0.022 & 0.809 & 0.850 & 0.925 \\
    MSE   & 0.014 & 0.005 & 0.006 & 0.014 & 0.029 \\
    RP    & $>0.999$     & 0     & $>0.999$     & $>0.999$     & $>0.999$ \\
    \hline
    \textbf{GSB} & Mean  & SD    & Min   & Median & Max \\
    Average Width & 1.112 & 0.022 & 1.069 & 1.108 & 1.171 \\
    MSE   & 0.016 & 0.005 & 0.008 & 0.015 & 0.030 \\
    RP    & 0.830 & 0.160 & 0.266 & 0.890 & $>0.999$ \\
        \hline
    \end{tabular}%
    \caption{Long range noise. Replicated results comparing the three spatial basis: (i) CHSB  (composite-hybrid spatial basis) with $\Upsilon=\ve\Phi\ve\Psi$; (ii) LSB (local spatial basis) with $\Upsilon=\ve\Phi$; (iii) GSB (global spatial basis) with $\Upsilon=\ve\Psi$, the principal components approach for spatial dependencies. The numbers indicate distribution summaries of the average joint bands interval width (Average Width) computed across all 25600 voxels, the mean squared errors (MSEs), and the proportion of real positives (RP) of the contrast estimator, computed over one hundred replications. }
  \label{table:sim2}%
\end{table}%

\subsection{Null simulation}

In this study we mimic resting states scenarios, situations where no task is being performed. The goal is to show that we are not detecting activation when it does not exist. For $t=1,\ldots,100$, $v$ indicates the voxel location in $\Re^{32 \times 32 \times 25}$, let
\begin{equation}
    Y_t(v)=\bar{Y}_C(v)+E_t(v),
\end{equation}
\noindent where $\bar{Y}_C(v)$ is the average taken over all time points of a true brain signal from the example subject described in Section \ref{sec:app}, and $E_t(v)$ follows the long-range spatial and autoregressive in time correlation structure previously described. Next, we use the same design matrix as in the previous example and estimate the contrast between Stimulus 1 and 2 for the following models: (i) CHSB (composite-hybrid spatial basis) with $\Upsilon=\ve\Phi\ve\Psi$; (ii) LSB ( local spatial basis) with $\Upsilon=\ve\Phi$; (iii) GSB ( global spatial basis) with $\Upsilon=\ve\Psi$, the principal components approach for the spatial domain; and (iv) NSB (no spatial basis) where we considered spatial independence across voxel. As before, we considered the temporal correlation as described in Section \ref{model:temporalbasis} and computed the mean squared error and false positive rate. For all compared models, we did not find any significant voxels based on the simultaneous credible bands (SimBas) presented in Section \ref{subsec:simbas}. 

As in the previous section, we replicated the comparison of the three spatial basis models one hundred times. In all one hundred cases we found no false-positives when using any spatial basis models. The average of the one hundred cases for the  mean squared error (computed across voxels) was 4.3036e-05 for the composite-hybrid spatial basis model, 3.8632e-05 for the local basis model, and 0.0026 for the global spatial basis model. 

The simulation results indicate that including a local basis based on spatial clusters (CHBM and LSB) increases statistical power in the joint bands to detect true differences in contrast estimation. This gain of power in the joint bands is not caused by an increase in the false positive rate. 

\section{Application}
\label{sec:app}

We analyzed the data from the Working Memory task of the Human Connectome Project (HCP). We considered the volumes collected from the right-left phase of the example Subject 100307. The experiment consists of a total of 8 blocks, and during each section,  volunteers  were shown a series of 10 images per block. Images corresponding to 4 stimuli (tools, places, body parts, faces) were embedded within the memory task. Each image was shown for 2.5s, followed
by a 15s inter-block interval. fMRI volumes were obtained every 720 ms. Each volume consisted of images of size $91 \times 109 \times 91$ for a total of 405 time frames. 

We partitioned the brain into clusters as described in Section \ref{sec:meth} using a digitalized version of the original Talairach structural labeling that was registered into the  MNI152 space. The atlas can be obtained from the FSL atlas library with information found at {\url{ https://fsl.fmrib.ox.ac.uk/fsl/fslwiki/Atlases}}  (\cite{Jenkinson2012}). We considered ROIs that had at least 125 voxels and obtained a total of 298 regions. Connectivity-based parcellation atlases have been proposed in the literature and could be used to guide the partitioning into clusters. However, there are some considerations to keep in mind. First, there is indication that functional parcel boundaries reconfigure with cognitive state (\cite{Salehi2020}). Second, they are usually focused on specific regions of the brain such as the  parietal cortex, the dorsal frontal cortex, and the cingulate and orbitofrontal cortex (\cite{Mars2011a,Sallet2013,Neubert2015}). In this application, we have chosen the structural Tailarach labeling because we wanted to focus on a more detailed brain parcellation to avoid missing unknown background connections. Nevertheless, we also provide an analysis with  the automated anatomical parcellation (AAL) in the supplementary material to confirm robustness of the proposed methodology.

\subsection{Estimation}

\begin{figure}
    \centering
    \includegraphics[width=0.5\textwidth]{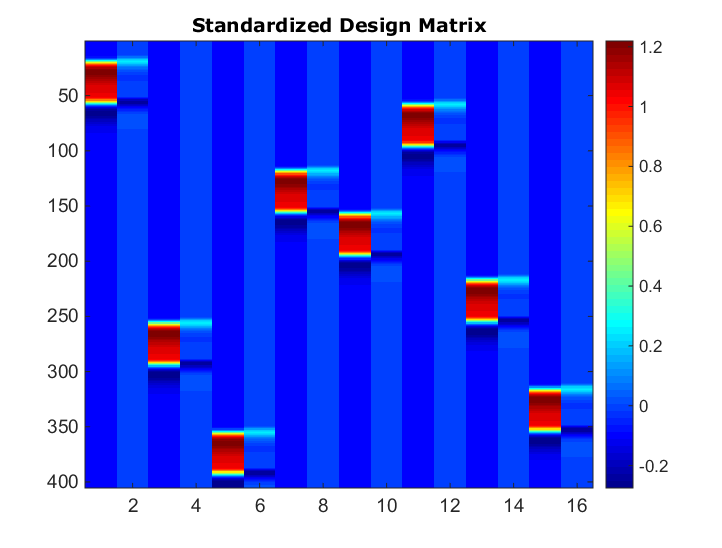}
    \caption{The design matrix X. Each odd column represents the convolution of a particular stimulus indicator and the haemodynamic response function and each even column represents the convolution of a particular stimulus indicator and the derivative of the haemodynamic response function. From left to right,  we have the following stimulus order:  2bk\_body, 2bk\_face, 2bk\_place,  2bk\_tool, 0bk\_body, 0bk\_face, 0bk\_place, and 0bk\_tool. }
    \label{fig: DesignMatrix}
\end{figure}

Let $Y_t(v)$ be the BOLD signal at frame $t = 1,\ldots, T = 405$, and voxel $v = 1,\ldots, V = 91 \times 109 \times 91 = 902,629$, then we assume

\[Y_t(v)=\sum_{a=1}^{16} X_{ta}B_{a}(v)+E_t(v),\]

\noindent where the odd columns of the design matrix X are formed by the convolution of each stimulus with the double-gamma HRF, and the even columns of X are formed by the convolution of each stimulus with the derivative of the double-gamma HRF. All columns are standardized to mean zero and variance one. The eight stimulus are: 2bk\_body, 2bk\_face, 2bk\_place,  2bk\_tool, 0bk\_body, 0bk\_face, 0bk\_place, and 0bk\_tool. Figure \ref{fig: DesignMatrix} shows a representation of the design matrix X, indicating when each stimulus happened during the experiment.

We estimate the coefficients $\ve B$ using five different models. The first four follow the methodology describe in Section \ref{sec:meth} with specific choices for the spatial basis and the same choice of temporal basis as detailed in Section \ref{model:temporalbasis}. The fifth model is the linear model from the software SPM (Statistical Parametric Mapping) with errors following an autoregressive process of order 1. The five models are:

\begin{enumerate}
    \item[(i)] Composite-hybrid spatial basis (CHSB) with $\Upsilon=\ve\Phi\ve\Psi$ and wavelet in time.
    \item[(ii)]  Local spatial basis (LSB) with $\Upsilon=\ve\Phi$ and wavelet in time.
    \item[(iii)] Global spatial basis (GSB) with $\Upsilon=\ve\Psi$, the principal components approach for the spatial domain, and wavelet in time.
    \item[(iv)] No spatial basis (NSB), where  we considered spatial independence across voxels, and wavelet in time.
    \item[(v)] SPM linear model with autoregressive process of order 1 in time.
\end{enumerate}

For models 1-4, we considered the long-term memory process detailed in Section \ref{model:temporalbasis} with 6000 MCMC samples, a burn-in of 2000 and thinning of 5.

\subsection{Results}
\label{sec:app:results}

We are interested in identifying brain activation patterns associated with different stimuli and, most importantly, estimating the background connectivity associated with the Working Memory task. We collapse across memory load and focus on the contrast  ``Faces versus Places''. This contrast is a good candidate because there is evidence that it will engage distinct cortical regions mainly on the fusiform face area and occipital gyrus (\cite{Barch2013}). 

\begin{figure}
    \centering
    \includegraphics[width=0.68\textwidth]{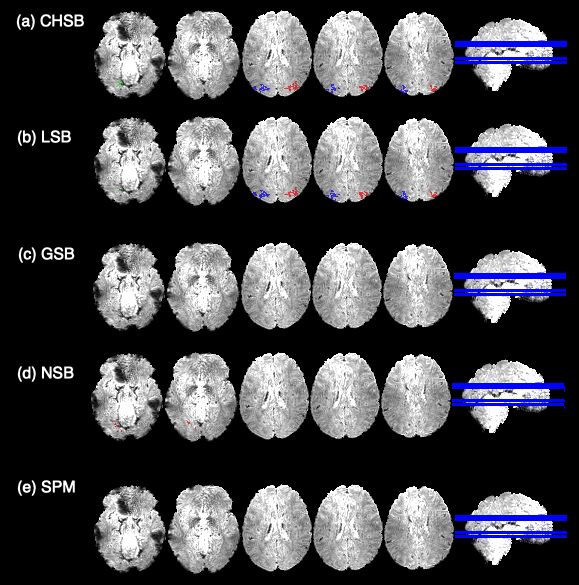}
    \caption{Axial slices of the significant location for ``Face vs Places'' for various models. Colors show clusters larger than 64 voxels. Panel (a) shows the significant locations using the proposed composite-hybrid model with local and global spatial features; panel (b) shows the significant locations using the composite-hybrid model with local features only; panel (c) shows the significant locations using the global features only; panel (d) shows the significant locations using the model with no basis; panel (e) shows the significant locations using the SPM software. }
    \label{fig:FacePlacesSign}
\end{figure}

We consider the contrast for Faces versus Places, $C_{FP}=1/2(B_3+B_{11})-1/2(B_5+B_{13})$, and evaluate whether the contrast in a particular voxel is significantly different than zero based on the joint bands SimBas as detailed in Section \ref{subsec:simbas}. The point estimate for each voxel, given by the posterior mean of the contrast $C_{FP}$, can be seen in Section 3 of the Supplementary Material. For SPM, the significance is giving by thresholding of the t-scores in each voxel. The threshold is obtained by the random field theory and it is given as an output by the software. Figure \ref{fig:FacePlacesSign} displays the locations in the brain where we flagged a difference based on the aforementioned joint bands. The different colors represent clusters of at least 64 voxels that share an edge in the 3D space. The numbers of voxels are shown in Table \ref{tableresults}. Columns 2-4 depict the number of clusters of significant voxels formed in each model, followed by the number of total voxels in all clusters in parenthesis. Clusters bigger than 125, 64, and 27 voxels are shown. Column 5 indicates the size of the biggest cluster, based on number of voxels, and Column 7 shows the average width of the jointband SimBas computed across all voxels.

\begin{footnotesize}
\begin{table}[]
\centering
\label{tableresults}
\begin{tabular}{|c|c|c|c|c|c|}
\hline
Face vs Places & \# Cl \textgreater{}125 & \# Cl \textgreater{}64 & \# Cl \textgreater{}27  & Bigg. Cl & Int. Width \\
\hline
 CHBM & 2 (652) & 2 (754) & 10 (1006) & 342 & 84.34 \\
 LSB & 2 (704) & 3 (783) & 8 (1003)  & 373 & 88.85 \\
GSB & 0 (0) & 0 (0) & 0 (0) & 8 & 146.60 \\
NSB & 1 (138) & 1 (138) & 6 (302)  & 138 & 99.26 \\
SPM & 0 (0)  & 0 (0) & 4 (143)  & 43 & ---\\
\hline
\end{tabular}
\caption{Columns 2-4 depict the number of clusters of significant voxels formed in each model. Clusters larger than 125, 64, and 27 voxels are shown, respectively. The number of clusters is followed in parentheses by the total number of voxels included in those clusters. Column 5 indicates the size of the biggest cluster in number of voxels, and Column 6 shows the average width of the joint bands computed across all voxels.}

\end{table}
\end{footnotesize}

\begin{figure}
    \centering
    \includegraphics[width=\textwidth]{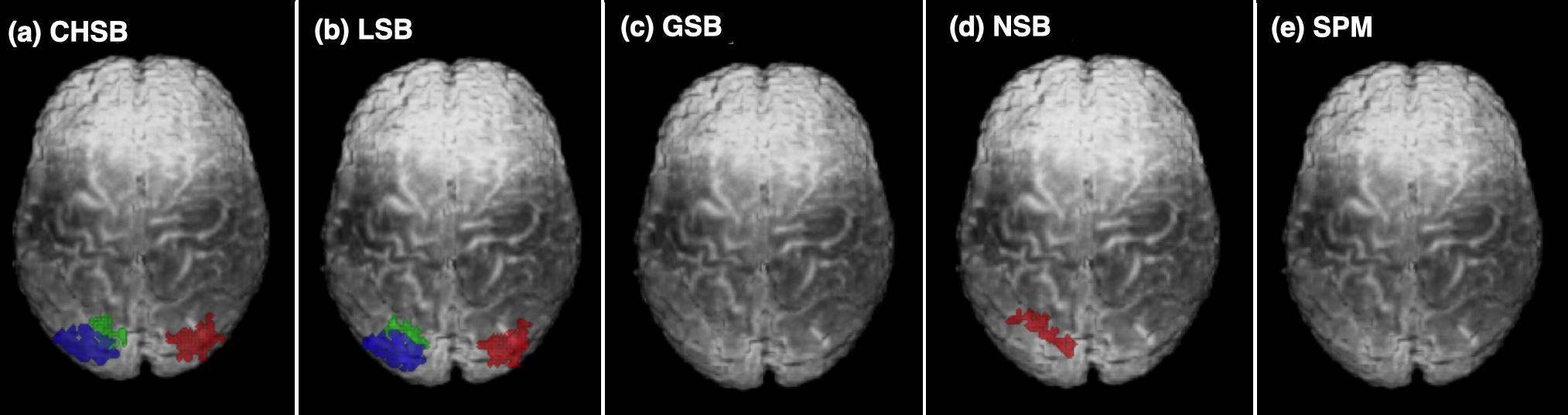}
    \caption{Differences between brain activity when processing images of faces versus places. Significant voxels are overlaid in a subject template. Panel (a) shows clusters larger than 64 voxels for the CHSB model, panel (b) shows clusters larger than 64 voxels for the LSB model, panel (c) shows clusters larger than 64 voxels for GSB panel, (d) shows clusters larger than 64 voxels for the NSB model, and panel (d) shows clusters larger than 64 voxels for the SPM model.}
    \label{fig:clusters}
\end{figure}

 Figure \ref{fig:clusters} is a render of Figure \ref{fig:FacePlacesSign} and displays the location of brain signatures after clustering;  only clusters larger than 64 voxels are shown. In Figure \ref{fig:FacePlacesSign}, the composite-hybrid model (panel a) indicates differences in the middle occipital gyrus (Regions 38 and 39); occipital cuneus (Region 486), and the posterior lobe declive in the cerebellum (Region 158). These regions are related to visual processing, face perception specialization, and cognition control, respectively (\cite{Pitcher2011,Park2018}).
 
 By jointly inspecting Figure \ref{fig:clusters} and Table \ref{tableresults}, we observe that accounting for local spatial structure considerably improves power to detect differences in the brain activation patterns (panels a and b).  Additionally, by inspecting panel (c) we conclude that naively computing a PC basis for the whole brain is not a good strategy. Panel (d) shows the results of the model with no spatial basis but long-memory assumptions in the temporal domain. The lack of spatial basis decreases power when compared to the models including a local spatial basis. However, its results are better than the SPM model with an AR(1) error assumption (panel e), indicating an adequacy of the long-memory assumption. As in the simulations, we note slightly tighter joint bands and more flagged regions in the CHBM including local and global spatial components than the LSB including just local spatial components, suggesting that not only did the CHBM yield estimates of background connectivity but its incorporation may have increased power for detecting activation signatures.


\begin{figure}
    \centering
    \includegraphics[width=\textwidth]{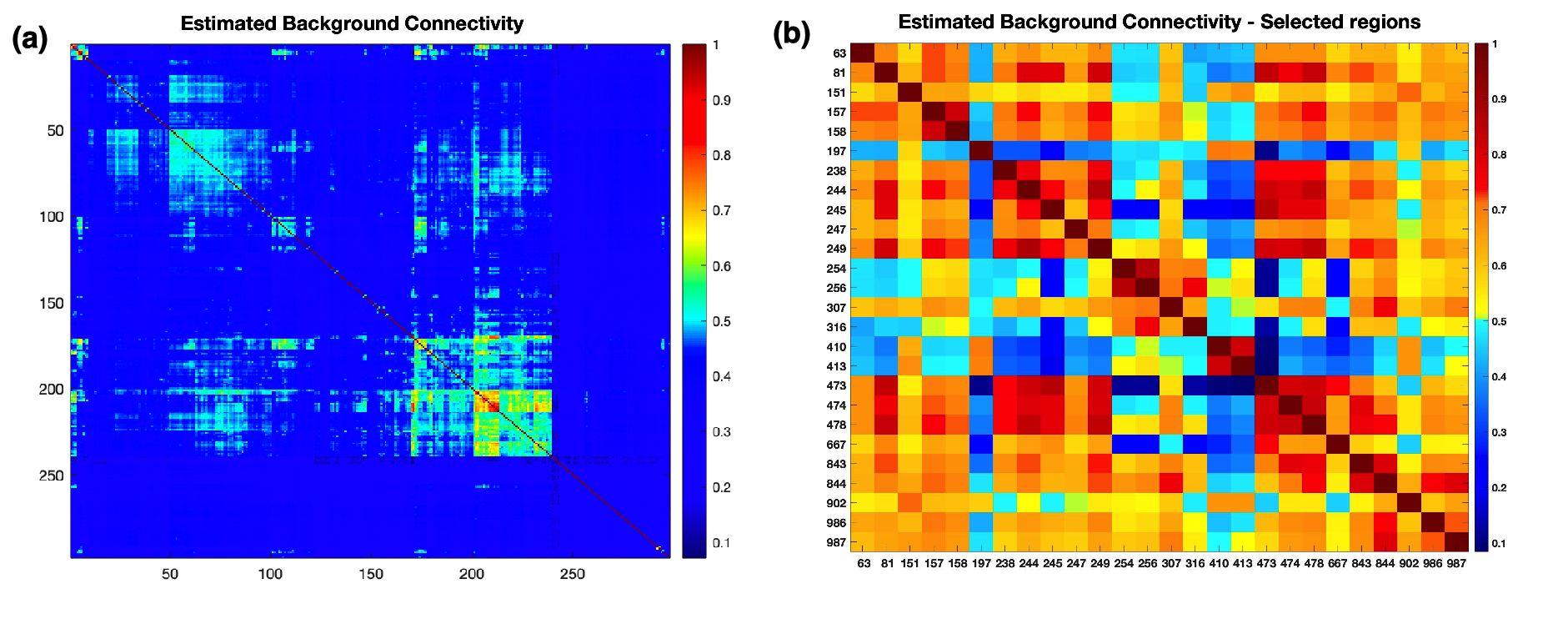}
    \caption{Estimated background connectivity. Values are the square root of the RV coefficients as defined in Section \ref{subsec:RV}. Panel (a) shows the estimated connectivity among the 298 ROIs. Panel (b) shows the estimated connectivity among the 26 selected regions. Regions were selected based on having at least one pair of connectivity higher than 0.7 and their labels are described in Table \ref{table:app:connec}.}
    \label{fig:resconn}
\end{figure}

We estimated the background connectivity among the 298 ROIs. Figure \ref{fig:resconn} panel (a) shows the results with values representing the square root of the RV coefficient detailed in Section \ref{subsec:RV}. These regions were ordered for better visualization and illustrate large connectivity among some ROIs.  Next, we selected regions that showed at least one connectivity value larger than 0.7 and find 26 unique ROIs. The connectivity among these regions is shown in Panel (b) of Figure \ref{fig:resconn}. The labelling numbers are the same as the original labelling of the Tailarach atlas and Table \ref{table:app:connec} has their corresponding description. We find strong connections among subregions in four distinct brain locations: frontal lobe, parietal lobe, limbic lobe, occipital lobe, and  cerebellum.

\begin{table}
\linespread{1.1}\selectfont\centering
\begin{footnotesize}
  \centering
    \begin{tabular}{|l|l|}
       
   \hline
       & {\bf Frontal lobe}\\
       \hline
   151&	Left Cerebrum.Frontal Lobe.Superior Frontal Gyrus.\\
    
    197	&Left Cerebrum.Frontal Lobe.Middle Frontal Gyrus.GM.BA 11\\
     410   & Left Cerebrum.Frontal Lobe.Middle Frontal Gyrus.GM.BA 10\\
    413   & Left Cerebrum.Frontal Lobe.Superior Frontal Gyrus.GM.BA 10   \\
    902	&Left Cerebrum.Frontal Lobe.Superior Frontal Gyrus.GM.BA 9\\
    \hline
       & {\bf Parietal lobe}\\
       \hline
    986	&Left Cerebrum.Parietal Lobe.Precuneus.GM.BA 19\\
    987   & Right Cerebrum.Parietal Lobe.Precuneus.GM.BA 19\\ 
    \hline
       & {\bf Limbic lobe}\\
       \hline
    667 &	Left Cerebrum.Limbic Lobe.Posterior Cingulate.\\
    \hline
       & {\bf Occipital lobe}\\
       \hline
    238   & Left Cerebrum.Occipital Lobe.Lingual Gyrus.  \\
    244   & Right Cerebrum.Occipital Lobe.Lingual Gyrus. \\
    245   & Right Cerebrum.Occipital Lobe.Lingual Gyrus.GM.BA 17  \\
    247	&Left Cerebrum.Occipital Lobe.Lingual Gyrus.GM.BA 18\\
    249   & Right Cerebrum.Occipital Lobe.Lingual Gyrus.GM.BA 18 \\
    254   & Right Cerebrum.Occipital Lobe.Inferior Occipital Gyrus.GM.BA 18  \\
    256   & Right Cerebrum.Occipital Lobe.Inferior Occipital Gyrus.WM.\\
     307   & Right Cerebrum.Occipital Lobe.Middle Occipital Gyrus.GM.BA 18  \\
     316&Right Cerebrum.Occipital Lobe.Middle Occipital Gyrus.GM.BA 19\\
     473   & Left Cerebrum.Occipital Lobe.Cuneus.  \\
     474   & Left Cerebrum.Occipital Lobe.Cuneus.GM.BA 18\\
    478   & Right Cerebrum.Occipital Lobe.Cuneus.GM.BA 18 \\   
    
    843	&Left Cerebrum.Occipital Lobe.Cuneus.GM.BA 19\\
    844   & Right Cerebrum.Occipital Lobe.Cuneus.GM.BA 19 \\
    \hline
       & {\bf Cerebellum}\\
       \hline
       63&	Left Cerebellum.Posterior Lobe.Uvula.GM.\\
   81    & Right Cerebellum.Posterior Lobe.Tuber.GM.  \\
   157   & Left Cerebellum.Posterior Lobe.Declive.GM.  \\
    158   & Right Cerebellum.Posterior Lobe.Declive.GM. \\
        \hline
    \end{tabular}%
    \caption{ROIs labellings for the 26 regions with background connectivity higher than 0.7 with at least one other region. GM and WM are gray matter and white matter, respectively. BA stands for Brodmann Area. A complete list can be found at http://www.talairach.org/labels.txt.}
  \label{table:app:connec}%
  \end{footnotesize}
\end{table}%

\begin{figure}
    \centering
    \includegraphics[width=\textwidth]{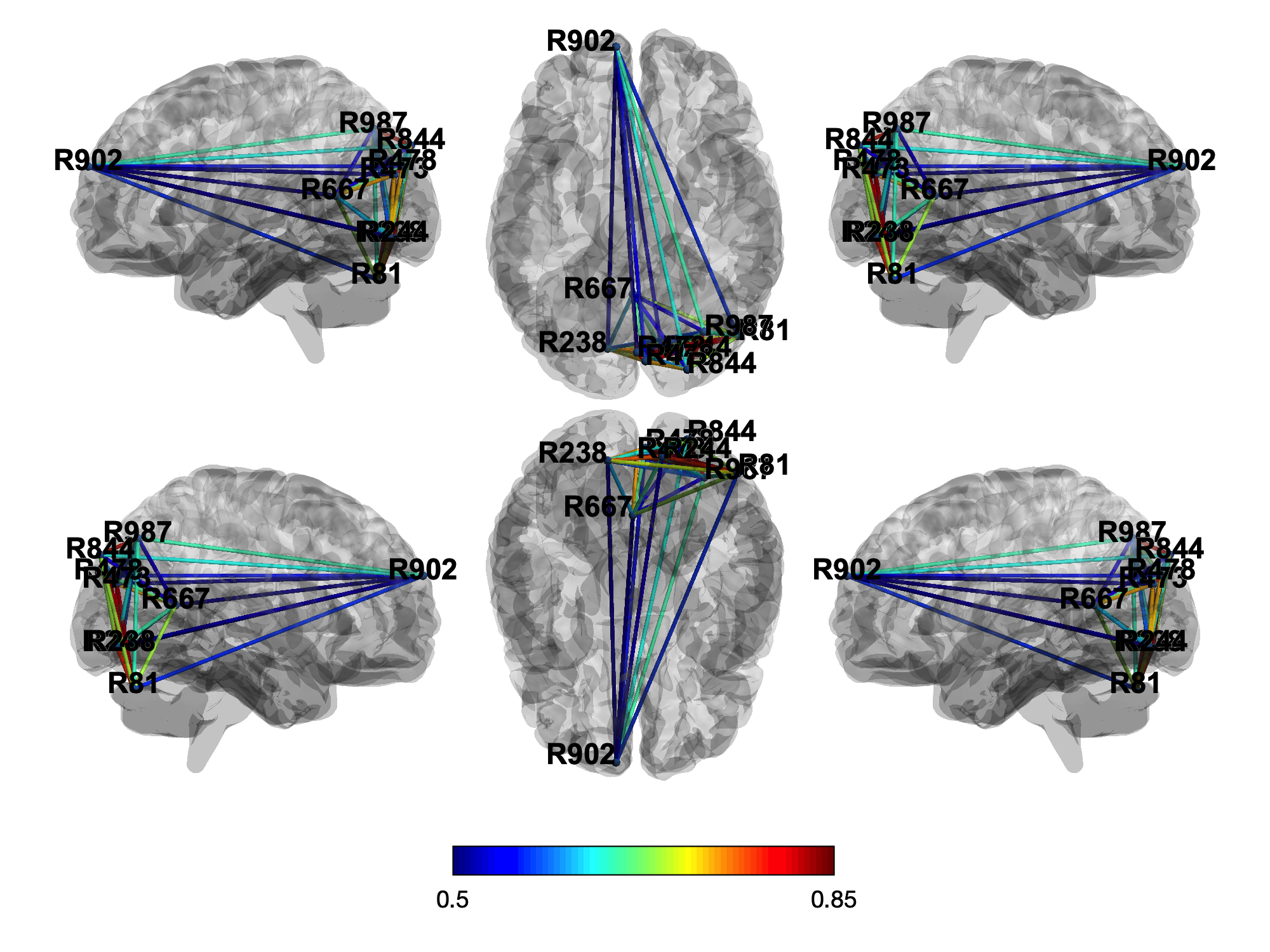}
    \caption{A subset of the estimated background connectivity network.  The edge colors vary according to the strength of background connectivity among the illustrated regions as indicated in the colorbar. The regions displayed are 81, 238, 244, 473, 478, 667, 844, 902, and 987 with label as in Table \ref{table:app:connec}.  We observe a complex network of areas specialized in vision, memory, and cognitive function with labels described in Table \ref{table:app:connec}.}
    \label{fig:resconn_network}
\end{figure}

For better visualization, we select a subset of the background connectivity values and illustrate the networks using BrainNet Viewer (\cite{Xia2013}). Figure \ref{fig:resconn_network} shows the results. The figure depicts connections among the five main regions previously mentioned.  The longer line in the figure represents a moderate connectivity between the frontal lobe (R902) and parietal lobe (R987). The node in the parietal lobe is in turn connected with the cuneus region in the occipital lobe (R844 and R478). The limbic lobe node is connected with multiple subregions in the occipital lobe and with the posterior lobe in the cerebellum (R81). High connectivity is observed between the cerebellum and occipital lobe indicating a complex cognitive network  engaged during the Working Memory task. 

These results complement the activation signatures found for the contrast Face versus Places. Although differences in activation patterns between images of faces versus places are captured in the occipital gyrus as expected, there is a complex functional network in place when executing the Working Memory task. There are many background connections to and from the cuneus  region (BA 18 in the occipital lobe), an area usually related to visual processing modulated by working memory.  We also observe a connection pattern  between the visual cortex (BA 17) and face perception specialization regions and those regions commonly engaged during complex cognitive tasks such as the frontal gyrus and the cerebellum (\cite{Boisgueheneu2016,Park2018}). The parietal lobe is connecting the frontal and occipital lobes as expected (\cite{Pitcher2011}), and connections of these regions with the limbic node are also observed. It is interesting to notice that the limbic node is often associated with memory. In summary, the task-dependent modulations that are due to variations in ongoing brain activity instead of stimulus-driven activity during the Working Memory task involve a complex network of areas specialized in vision, memory, and cognitive function.

Finally, we look at the computational time needed for computing initial values, estimating model parameters, projecting back to the data space, and calculating the joint bands for inference. Table \ref{table:apptime} shows the times in minutes for each component of the estimation procedure. This analysis was performed in a 2.8 GHz Intel Core i7 MacBook Pro with 16 GB of memory. It took less than 13 minutes to perform full Bayesian inference at the voxel and ROI level on a single processor.  Given cluster computing resources, this makes it feasible to perform many single-subject analyses in parallel.  
 
\begin{table}
\label{table:apptime}
\centering
\begin{tabular}{|l|r|r|r|}
\hline
Computational time (in min) & CHSB &LSB & GSB \\
\hline
Initial Values & 3.05 & 116.49 & 2.94\\
MCMC & 6.72 & 405.79 & 6.45\\
Projection & 2.58 & 2.58 & 0.05\\
SimBas & 0.08 & 0.08 & 0.08\\
\hline
\end{tabular}
\caption{Total computational time (in minutes) of the full Bayesian MCMC estimation for the composite-hybrid  model separated by task. {\bfseries Variance Components} indicates the total time to estimate the initial values of the variance components of the long-memory prior.  {\bfseries MCMC} indicates the total time to run the MCMC algorithm with 6000 iterations. {\bfseries Projection} indicates the total time to project the estimated values into the original voxel space, and {\bfseries SimBas} indicate the total time to calculate p-values and simultaneous bands that control for the experiment-wise errors as described in Section \ref{subsec:simbas}.}
\end{table}

\section{Discussion}
\label{sec:discus}

We have proposed new Bayesian methodology for joint estimation of activation signatures and background connectivity patterns. The proposed model combines strength of individual basis approaches by choosing a  basis  set that is composite-hybrid. This set  effectively  provides  a  tensor  time-space  basis  representation  of   the temporally-varying volumetric 4d task fMRI data while flexibly capturing  key characteristics of the data. Additionally, the composite-hybrid approach is adaptive yet provides a strong degree of dimensionality reduction and regularization enabling it to scale up to large fMRI data sets.

The composite-hybrid has greater power compared to a variety of alternative strategies because it incorporates local information through clusters, while accounting for possible distant correlations between distant brain regions. In turn, the composite-hybrid approach provides a fast algorithm that can be used in parallel to estimate the activation signatures and background connectivity of multiple subjects.  This would make feasible a rigorous full Bayesian group-level analyses in just a couple of hours.  We have devised a strategy for group-level analysis for activation signatures and background connectivity that aggregates posterior predictive distributions of the subject specific models, propagating their uncertainty, into a group-level model in which we will learn group level hybrid basis functions from the individual basis and construct a novel projection-based approach.  This requires substantial additional development, so we leave this to future work, but the speed and flexibility of the subject-level analyses presented in this paper is key to making that strategy possible.


Finally, looking at background connectivity is a step forward in fully understanding ongoing brain activities and should be used to complement task-induced  and resting state connectivity studies. The joint fully Bayesian background connectivity estimation method is a powerful tool that will allow discovery of compelling insights into brain networks from task fMRI data.






\section*{Funding}
Jeffrey Morris is supported by grants R01-CA-178744 and R01-CA-244845 from the National Cancer Institute.
\bibliographystyle{plain}

\bibliography{HCP_References}

\end{document}